# Limitations to THz generation by optical rectification using tilted pulse fronts

**Koustuban Ravi [1*], W.Ronny Huang[1], Sergio Carbajo[2,3], Xiaojun Wu[2] and Franz Kärtner[1,2,3]**

[1] Department of Electrical Engineering and Computer Science, Research Laboratory of Electronics, Massachusetts Institute of Technology, Cambridge, MA 02139, USA
[2] Center for Free-Electron Laser Science, Deutsches Elektronen Synchrotron, Hamburg 22607, Germany
[3] Department of Physics and the Hamburg Center for Ultrafast Imaging, University of Hamburg, Germany

[*]koust@mit.edu

**Abstract:** Terahertz (THz) generation by optical rectification (OR) using tilted-pulse-fronts is studied. One-dimensional (1-D) and 2-D spatial models, which simultaneously account for (i) the nonlinear coupled interaction of the THz and optical radiation, (ii) angular and material dispersion, (iii) absorption, iv) self-phase modulation and (v) stimulated Raman scattering are presented. We numerically show that the large experimentally observed cascaded frequency down-shift and spectral broadening (cascading effects) of the optical pump pulse is a direct consequence of THz generation. In the presence of this large spectral broadening, the phase mismatch due to angular dispersion is greatly enhanced. Consequently, this cascading effect in conjunction with angular dispersion is shown to be the strongest limitation to THz generation in lithium niobate for pumping at 1μm. It is seen that the exclusion of these cascading effects in modeling OR, leads to a significant overestimation of the optical-to-THz conversion efficiency. The simulation results are supported by experiments.

**1. Introduction**

Terahertz (THz) radiation with high pulse energies is of significant interest to many applications such as non-linear spectroscopy [1], charged particle acceleration [2]-[3], high harmonic generation [4] and molecular alignment [5], to name a few. Of various THz generation modalities, optical rectification (OR) of femtosecond laser pulses with tilted-pulse-fronts in lithium niobate has emerged as the most efficient THz generation technique. This approach [6] produces single-cycle THz fields with optical-to-THz conversion efficiencies (henceforth referred to as conversion efficiency) in excess of 1% at room temperature [7]. Consequently, the approach has attracted a lot of interest in the pursuit of mJ-level THz pulse energies [8]. However, the theoretically predicted conversion efficiencies for this approach are larger than those of the corresponding experimental demonstrations. For instance, conversion efficiencies of 10% at 100K are predicted for transform limited 500 fs full-width-at-half-maximum (FWHM) pulses centered at a wavelength of ~1μm [8]-[9]. On the other hand, experimental results [7,9,10] have not yielded similar values. The disparity between experimental and theoretical results motivates a re-examination of existing models to identify potential shortcomings in theory.

In [11], a one-dimensional (1-D) spatial model including the effects of material dispersion and group velocity dispersion due to angular dispersion (GVD-AD) was presented. A Drude model for free-carrier-absorption (FCA) of THz radiation was used to model the saturation of THz energy. In [12] along with experiments, a 1-D model considering material dispersion, GVD-AD and self-phase modulation (SPM) suggested that SPM rather than FCA was the principal reason for saturation of THz energy. In [12]-[13], a 2-D model which took into account material dispersion, GVD-AD and crystal geometry was developed.

In the aforementioned models [11- 13], the spectral re-shaping of the optical pump pulse caused by THz generation was not considered. However, these effects have to be included to explain the dramatic experimentally observed frequency down-shift and spectral broadening of the optical pump pulse, commensurate with the amount of THz generated using both 1µm [7] as well as 800 nm pump sources [14]. This spectral broadening can intuitively be understood to be a consequence of a cascaded down-shift of the optical pump frequency upon generating THz photons. We will henceforth refer to the cascaded down-shift and spectral broadening as 'cascading effects'.

In [15], a 1-D model which accounted for these cascading effects was presented. In [16], we extended this model to account for GVD-AD and material dispersion in tilted-pulse-fronts. In this paper, we present 1-D and 2-D spatial models which simultaneously account for (i) cascading effects, (ii) material and angular dispersion, (iii) absorption, (iv) SPM and (v) stimulated Raman scattering (SRS). The presented models are applicable to simulating a wide variety of OR devices based on different material systems and crystal geometries.

Using the 1-D model, we are able to confirm that the experimentally observed cascading effects are a direct consequence of THz generation. The extent of spectral broadening is demonstrated to be commensurate to the amount of THz generated. Furthermore, in both experiments and theory, it is seen that SPM effects cause negligible spectral broadening. The quantitative agreement of simulated optical spectra and conversion efficiency with experiments [7] is a good validation of the model.

Due to the large spectral broadening resulting from THz generation, the phase mismatch due to GVD-AD is significantly accentuated. Consequently, the conversion efficiency decays much more drastically with propagation length than simply with other limiting effects such as SPM. Therefore, cascading effects in conjunction with GVD-AD represent the strongest limitation to achieving high conversion efficiencies in lithium niobate, especially for pumping at 1μm (or longer wavelengths). For pumping at 800nm, in addition to the limitations imposed by cascading effects, FCA may also be an impediment to THz generation.

We show that neglecting these cascading effects, i.e. the spectral re-shaping of the optical pump spectrum, can lead to a significant overestimation of achievable conversion efficiencies. For example, when cascading effects are excluded, conversion efficiencies as high as 10% have been predicted [8]. However, for similar conditions, the consideration of cascading effects results in conversion efficiencies of only ~2%.

Since, the accurate modeling of OR of an angularly dispersed femtosecond pulse with a tilted-pulse-front is inherently a 2-D problem, we develop a comprehensive 2-D spatial model that solves simultaneously for both optical and THz electric fields. This model confirms that cascading effects in conjunction with GVD-AD represent the strongest limitation to THz generation in lithium niobate for 1μm pumping.

Since THz generation will produce cascading effects in any OR process, the phase mismatch due to dispersion (angular or material) will be enhanced. Consequently, it would be important to account for them in a realistic analysis of any OR system.

**2. Description of Optical Rectification using tilted pulse fronts**

In lithium niobate, the refractive indices at THz (~5) and optical frequencies (~2.2) are highly mismatched. Consequently, broadband phase matching using collinear geometries in bulk lithium niobate is not possible. In order to circumvent this problem, it was proposed that an angularly dispersed optical pump pulse be used [6]. The angular dispersion is introduced via a diffraction grating as shown in Fig.1. The optical pump pulse is incident at an angle $\theta_i$ with respect to the normal of the grating surface and each frequency component of the optical pump pulse with angular frequency ω scatters off at a different angle $\theta_d(\omega)$. A focusing lens, located at a distance $s_1$ from the diffracting grating is used to image this angularly dispersed optical pump pulse onto the lithium niobate crystal located at a distance $s_2$ from the lens as shown in Fig.1a. In the angularly dispersed pulse, each frequency at ω, has a corresponding propagation vector $\vec{k}(\omega)$. Two such components are depicted schematically in Fig.1. Since OR is essentially difference frequency generation (DFG), for phase matching to occur, the propagation vector of the generated THz (at angular frequency Ω) is given by $\vec{k}(\Omega) = \vec{k}(\omega+\Omega) - \vec{k}(\omega)$ as shown in label (i) in Fig.1. The generated THz then emerges at an angle γ with respect to the direction of propagation of the optical pump pulse. The apex angle of the crystal in such systems is designed to be ~γ so that the THz exits normal to the output facet of the crystal. It turns out that, the spatial intensity profile of the angularly dispersed optical pump pulse at any given time instant is tilted w.r.t its propagation direction by an angle π/2- γ. It is important to observe that the tilted intensity front is not the same as the phase front. This, tilted intensity profile is defined as a 'tilted-pulse-front' with pulse-front-tilt angle γ. THz propagation occurs perpendicular to this tilted-pulse-front, which is once again at an angle γ with respect to the direction of optical pump propagation. Based on Fig.1, we see that if the THz propagates a distance z, in the phase matched direction, the optical pump pulse propagates a distance z/cosγ due to its greater velocity (or smaller refractive index).

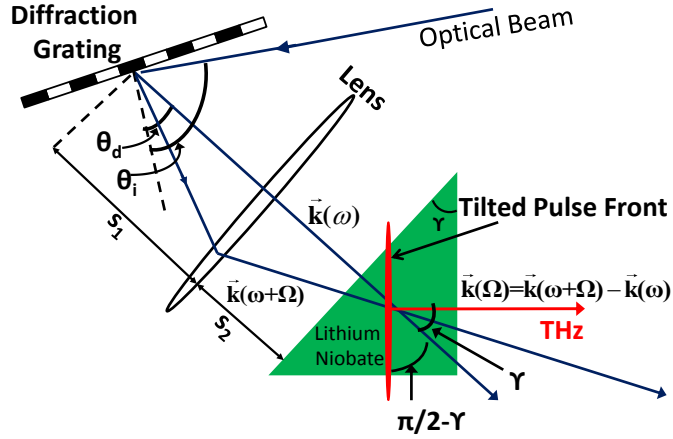

Fig.1. The setup for OR using tilted pulse fronts. An optical pump pulse is incident at an angle θ$_i$ onto a diffraction grating. Frequency components are scattered off at different angles $\theta_d(\omega)$ and are imaged onto the lithium niobate crystal by a focusing lens. Phase matching occurs as in DFG and the THz propagates at an angle γ with respect to the optical pump. The spatial profile of the angularly dispersed beam at any time instant is tilted with respect to its propagation direction and is called the tilted-pulse-front, with tilt angle defined as γ. If the THz wave propagates a distance z, the optical wave has propagated a distance z/cosγ.

## 3. Theory

We first explore the impact of cascading effects using an effective 1-D spatial model. The electric fields corresponding to the optical pulse at spatial co-ordinate z and various angular frequencies ω are denoted by $E_{op}(\omega,z) = A_{op}(\omega,z)e^{-jk(\omega)z}$. The THz electric field at spatial co-ordinate z and angular frequency Ω is denoted by $E_{THz}(\Omega,z) = A_{THz}(\Omega,z)e^{-jk(\Omega)z}$. Here, $A_{op}(\omega,z)$ and $A_{THz}(\Omega,z)$ represent the envelope of the optical and THz fields (henceforth referred to as field) respectively. The wave number for the THz field is given by $k(\Omega) = \Omega n(\Omega)/c$, where c is the speed of light in vacuum. Exact values of the THz refractive index $n(\Omega)$ are used to accurately model the material dispersion in the THz range. The wave number for the optical field is given by Eq. (1a).

$$k(\omega) = \frac{1}{\cos\gamma}\frac{\omega n(\omega)}{c} + \frac{(\omega-\omega_0)^2}{2}k"(\omega)_{AD} \qquad (1a)$$

$$k"(\omega)_{AD} = \frac{-n_g^2(\omega_0)}{\omega_0 c n(\omega_0)}\tan^2\gamma \qquad (1b)$$

Here, the refractive index $n(\omega)$ accounts for material dispersion. The factor $\cos\gamma$ appears in the denominator of the first term in Eq. (1a) to account for the tilted-pulse-front, where γ is the pulse-front-tilt angle. The second term $k"(\omega)_{AD}$, corresponds to GVD-AD. The expression for GVD-AD is given by Eq.(1b). In Eq.(1b), $\omega_0$ is the angular frequency at which the pulse spectrum of the optical pulse is initially centered. The parameter $n_g(\omega_0)$ is the group refractive index of the material at $\omega_0$. Note, that higher order expansion terms can also be derived in Eq.(1b) but it turns out that they minimally impact the results. When the tilt angle γ=0, Eq.(1a) reduces to the regular expression for the wave number. Therefore, the developed model can be applied to both non-collinear and collinear OR geometries. It is worthwhile to note that the exact material dispersion and GVD-AD were not considered in [15], which underestimated their detrimental effects. The evolution of $A_{THz}(\Omega,z)$ is given below in Eq.(2).

$$\frac{dA_{THz}(\Omega,z)}{dz} = -\frac{\alpha(\Omega)}{2}A_{THz}(\Omega,z)$$
$$-\frac{j\Omega^2}{2c^2 k(\Omega)}\chi_{eff}^{(2)}(z)\int_0^\infty A_{op}(\omega+\Omega,z)A_{op}*(\omega,z)e^{-j[k(\omega+\Omega)-k(\omega)-k(\Omega)]z}d\omega \qquad (2)$$

Here, the first term on the right hand side (RHS) corresponds to the THz loss due to absorption, which may include background as well as FCA. The second term corresponds to the second order non-linear polarization due to OR. Noticing this term, it is clear that OR is nothing but an aggregate of all possible difference frequency generation (DFG) processes within the bandwidth of the optical pulse. It can be seen that setting the exponential factor inside the integral on the RHS of (2) to zero, gives the well-known phase matching condition $n(\Omega)\cos\gamma = n_g(\omega_0)$ for OR using tilted-pulse-fronts. In Eq.(2), the parameter $\chi_{eff}^{(2)}(z) = 2d_{eff}(z)$ is the effective second order non-linear susceptibility at the spatial co-ordinate z. For the calculations in this paper, $\chi_{eff}^{(2)}(z)$ is constant in space. However, by introducing spatial variations in $\chi_{eff}^{(2)}(z)$, the model can also be used to analyze quasi-phase-matching (QPM) structures. Thus, the developed model can be used to simulate a number of devices based on OR. If only Eqs.(1)-(2) are solved, we would be working in the undepleted pump approximation.

In order to account for pump depletion and spectral re-shaping of the optical pump, we write the corresponding equations for the evolution of $A_{op}(\omega,z)$ in Eq. (3).

$$\frac{dA_{op}(\omega,z)}{dz} = \frac{-\alpha_{op}A_{op}(\omega,z)}{2}$$
$$-\frac{j\omega^2}{2c^2k(\omega)}\chi_{eff}^{(2)}(z)\int_0^\infty A_{op}(\omega+\Omega,z)A_{THz}^*(\Omega,z)e^{-j[k(\omega+\Omega)-k(\omega)-k(\Omega)]z}d\Omega$$
$$-\frac{j\omega^2}{2c^2k(\omega)}\chi_{eff}^{(2)}(z)\int_0^\infty A_{op}(\omega-\Omega,z)A_{THz}(\Omega,z)e^{-j[k(\omega-\Omega)-k(\omega)+k(\Omega)]z}d\Omega$$
$$+\mathbf{F}\left\{j\frac{\varepsilon_0\omega_0 n(\omega_0)^2 n_2(z)}{2}|A_{op}(z,t)|^2 A_{op}(z,t)\right\} + \mathbf{F}\left\{j\frac{\varepsilon_0\omega_0 n(\omega_0)^2 n_2(z)}{2}\left[|A_{op}(z,t-t')|^2 \otimes h_r(t')\right]A_{op}(z,t)\right\}$$

(3)

In Eq. (3), the first term corresponds to the absorption of the optical radiation. The second term is analogous to the second term in Eq. (2), and represents the cascaded down-conversion of the optical frequency due to THz generation. The third term represents the up-conversion of optical frequency components due to sum frequency generation (SFG) between an optical frequency component at (ω-Ω) and a THz field component at Ω. The fourth term corresponds to SPM. Here $\varepsilon_0$ is the permittivity of free space and $n_2(z)$ is the non-linear refractive index coefficient at spatial co-ordinate z. The symbol **F** is used to denote the temporal Fourier transform. It can be seen that the effects of SPM are directly proportional to the optical field intensity. The final term corresponds to SRS. Here $h_R(\omega')$ represents the Raman gain spectrum. The complex second order nonlinear polarization terms on the right hand side of Eqs. (2)-(3), were calculated rapidly by expressing them in terms of Fourier transforms. A 4$^{th}$ order Runge-Kutta method was used to numerically solve the above equations with a spatial resolution of 10μm. This spatial resolution was verified to achieve numerical convergence.

## 4. Validation of model: Comparison to experimentally measured optical spectra

In order to validate the developed 1-D model, we simulate the transmitted spectrum of the optical pump pulse at various levels of THz generation for the experimental parameters presented in [7]. The optical pump pulse used in [7] was modeled by a transform limited 500 fs Gaussian pulse centered at 1031.8 nm with a maximum fluence of 20mJ/cm$^2$. The second order effective non-linear susceptibility $\chi_{eff}^{(2)} = 2d_{eff}$ was assumed to be $360\,pm/V$ [18]. A non-linear refractive index value of $n_2 = 1 \times 10^{-15}\,cm^2/W$ [19] was used to account for SPM and SRS effects. The Raman gain spectrum was obtained from measurements in [20]. The refractive index and absorption coefficient data at optical and THz frequencies were obtained from [21]. At 1µm pump wavelengths, FCA of THz radiation occurs mainly due to free-carriers generated by four-photon absorption of the optical radiation. Experimental fits have estimated a four-photon absorption coefficient of 10$^{-7}$cm$^5$/GW$^3$ [22] and a THz FCA cross section of $2 \times 10^{-21}\,cm^2$ [22]. Corresponding to these values, even a 500fs pulse with peak intensity of 150 GW/cm$^2$ will result in a FCA coefficient of only ~0.1cm$^{-1}$, which is much smaller than the background absorption coefficient of lithium niobate (~10cm$^{-1}$ at room temperature [21]). Therefore, for the peak intensities of 40GW/cm$^2$ used in the simulations below, FCA and four-photon absorption coefficients can be effectively set to zero with negligible loss in accuracy. The crystal temperature was assumed to be 290K. The pulse-front-tilt angle for optimal phase matching at 290K was γ=63.09°. A Fresnel reflection loss of ~44% due to the high refractive index of lithium niobate at THz frequencies is included in the calculations.

In the actual physical situation, different parts of the optical pump beam see different propagation lengths due to the prism geometry of the crystal in OR using tilted-pulse-fronts. This situation can be modeled accurately only with a 2-D spatial model, as will be done in Section 6. However, in the 1-D model, an effective propagation length L$_{eff}$ is used. The simulated spectra of the transmitted optical pulses are plotted in Fig.2 along with the experimental results. The conversion efficiency is tuned by changing the pump fluence from zero to a maximum of 20mJ/cm$^2$. This mimics the rotation of polarization in the actual experiment. The maximum conversion efficiency at room temperature was reported to be 1.15% in [7]. We obtain a maximum value of 0.8% in our simulations, which is in good agreement with the experimental results. The difference in results can be attributed to small uncertainties in material parameters ($\chi_{eff}^{(2)}, n_2, \alpha_{THz}$) and experiments. In particular, there is uncertainty in the absorption coefficients below 0.9THz [21].

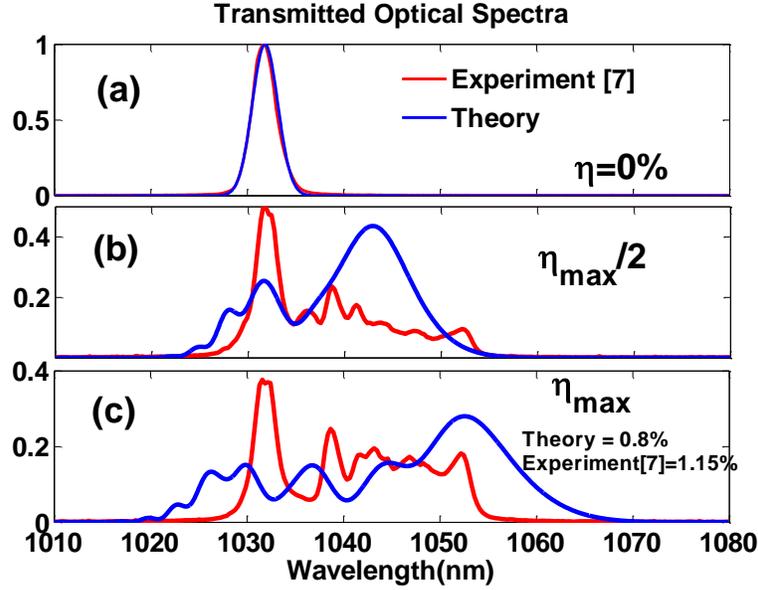

Fig.2. Comparison of experimental and simulated optical spectra for different amounts of generated THz. The frequency down-shift and spectral broadening can be modeled only by the simultaneous solution of the THz and optical fields. (a) No broadening or red-shift is observed when conversion efficiency η=0 which implies small effect of SPM. (b)There is large cascaded frequency down-shift and spectral broadening corresponding to a larger amount of THz generation (c) Maximum frequency down-shift and spectral broadening is observed when conversion efficiency is maximum at $\eta_{max}$=0.8%. Conversion efficiency, amount of frequency down-shift and spectral broadening are in good agreement with experiments [7]. The difference between theory and experiment can be attributed to small uncertainties in material parameters ( $\chi^{(2)}_{eff}, n_2, \alpha_{THz}$ ) and experiments. The measured spectra include spatial averaging effects which are not included in the calculations, which could explain their difference.

In Figs.2a-2c, both the simulated and experimental spectra show an amount of broadening and red-shift commensurate with the amount of THz generated. This is strong evidence that the observed broadening is a direct consequence of THz generation. To further reinforce this point, in Fig.2a, when there is virtually no THz generation, there is also negligible broadening of the transmitted optical pump pulse. This also indicates a relatively small impact of SPM and SRS effects in these experiments even at the relatively large peak intensities of 40GW/cm$^2$.

In Figs.2b and 2c, the extent of broadening seen in experiments is well reproduced by the simulations at half the maximum and maximum conversion efficiency, respectively. Here, in addition to the red shift, a relatively small amount of blue-shift is also seen which also increases with increasing THz generation. This effect is also observed in our calculations. An explanation for this is obtained by inspecting the third term on the right hand side of Eq. (3). This term represents a blue-shift of the optical pulse via THz plus optical SFG, which increases with the THz conversion. Consequently, there is an increasing amount of blue-shift with increased THz generation, albeit to a much lesser extent than that of the red-shift. The difference in the shape of the spectra between experiments and simulations is partly due to the usage of a 1-D model. In the actual physical situation, due to the non-collinear propagation of the optical and THz radiation, different parts of the optical beam will be broadened to different extents. The final recorded power spectrum would then correspond to a spatial averaging of the spectral intensity over the beam cross-section. Such spatial averaging has not been considered here.

The good match in conversion efficiency, frequency shifting, and spectral broadening between simulation and experiments is evidence that even this simple 1-D model captures the essential physics of the nonlinear THz generation process.

**5. Limitation of conversion efficiency due to cascading effects**

In Section 4, we established the accuracy of our model by obtaining good quantitative agreement with experiments in conversion efficiency values and transmitted optical spectra. We now use the model to show that cascading effects in conjunction with GVD-AD represent the strongest limitation to THz generation for 1µm pumping. We selectively switch-on and switch-off, various effects in our simulations to study their relative importance.

In Fig.3, we calculate the conversion efficiency as a function of the effective propagation length- $L_{eff}$ for the following cases :-(i) only GVD-AD is considered. This corresponds to an undepleted pump approximation where only Eqs.(1),(2) are used. There is no change in the total pump pulse energy or spectrum; (ii) SPM, SRS and GVD-AD are considered. Equations (1) and (2) are used. Only the fourth and fifth terms in Eq.(3) are employed. Here, the total pump pulse energy is conserved but spectral re-shaping of the pump pulse spectrum occurs due to SPM; (iii) cascading effects are included but SPM, SRS and GVD-AD are excluded. Equations (1a), (2) and the first three terms of Eq.(3) are utilized. The second term of Eq.(1a), $k''(\omega)_{AD}$ is set to zero ; (iv) cascading effects and GVD-AD are included but SPM and SRS are excluded. Equations (1)-(3) are used but the fourth and fifth terms of Eq.(3) are ignored;(v) all effects are simultaneously included (Eqs.(1)-(3)). Material dispersion and THz absorption are considered in all cases. A crystal temperature of 100K is assumed. This is because the difference between various cases is best illustrated with cryogenic cooling rather than at room temperature. The pump fluence is fixed at 20mJ/cm$^2$ for all simulations as in Section 4. The material parameters are the same as in Section 4, except for absorption and refractive index which are functions of crystal temperature. Due to the change in refractive index with temperature, the pulse-front-tilt angle for optimal phase matching is now set to γ=62.26°.

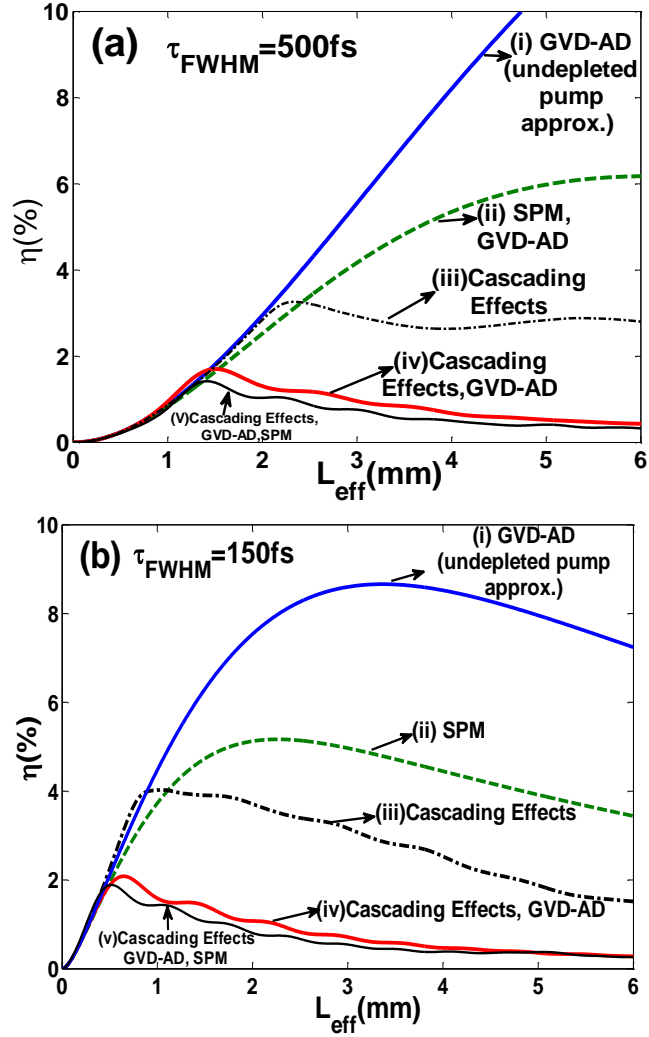

Fig.3: Conversion efficiencies as a function of effective length are calculated by switching on and switching off various effects. Material dispersion and absorption are considered for all cases. The pump fluence is 20mJ/cm$^2$, $\chi^{(2)}_{eff} = 360\,pm/V$, $n_2=10^{-15}$ cm$^2$/W and crystal temperature is 100K.(a) Gaussian pulses with 500 fs pulse width at FWHM with peak intensity of 40GW/cm$^2$ are used. Cascading effects together with GVD-AD leads to the lowest conversion efficiencies. The drop in conversion efficiency is attributed to the enhancement of phase mismatch caused by angular dispersion (GVD-AD) due to the large spectral broadening caused by THz generation (See Figs.2(b)-(c)). SPM effects are much less detrimental since they cause relatively small broadening of the optical pump spectrum (see Fig.2(a)). (b) Cascading effects along with GVD-AD are most detrimental even for a 150fs Gaussian pulse three times larger peak intensity.

In Fig.3a, we plot the simulation results for cases (i)-(v) for a transform limited Gaussian pulse with a pulse width of 500 fs at FWHM. This corresponds to a peak intensity of 40GW/cm$^2$. For case (i), when only GVD-AD is considered, conversion efficiencies larger than 10% are possible, in agreement with the results from [8]. For case (ii), when SPM and SRS effects are included, the conversion efficiency peaks at about 6%. The reduction in maximum conversion efficiency is because of the additional phase mismatch induced by the change in refractive index at optical frequencies due to SPM effects. The increase in phase mismatch is evident in the reduction of the optimal effective length value compared to case (i) to $L_{eff}$ = 6mm. Beyond this distance, the coherent build up of THz energy reduces as the optical and THz pulses slip out of phase.

In case (iii), when only cascading effects are included, the conversion efficiency peaks at ~3%. This shows that cascading effects are even more detrimental to THz generation than SPM. This is demonstrated by a further decrease of the optimal effective length to ~2.5mm. This trend agrees with the experiments and simulations in Fig.2. Here, it was seen that the spectral broadening which occurs as a consequence of THz generation was much larger than that due to SPM. Dispersive effects are expected to be enhanced for larger spectral bandwidths and therefore their corresponding optimal effective lengths are expected to be shorter.

In case (iv), when both cascading effects and GVD-AD are considered, the conversion efficiency peaks at 1.8%. The optimal effective length reduces even further to ~1.5mm, since GVD-AD is larger than material dispersion in lithium niobate. Finally, if all effects (SPM, GVD-AD, cascading effects) are considered simultaneously in case (v), there is only a minor change compared to case (iv) in Fig.3a. This confirms that the combination of spectral broadening caused by the generation of THz (cascading effects) and angular dispersion are the dominant limiting effects. A five-fold reduction in the maximum achievable conversion efficiency in comparison to the 10% efficiency levels for similar pump parameters [8] is observed.

In Fig.3b, we show the results for cases (i)-(v), using a transform limited Gaussian pulse with a pulsewidth of 150 fs at FWHM, while keeping the pump fluence constant. This corresponds to a peak intensity of ~120GW/cm$^2$. At these higher intensities, the effect of SPM is seen to be more pronounced than in Fig.3a. However, even here, cascading effects in conjunction with GVD-AD impose the strongest limitation to conversion efficiency. The effective lengths at which conversion efficiency peaks reduces even further compared to Fig.3a for all cases. In case (i), this is because of the larger effects of dispersion due to an increased spectral bandwidth compared to Fig.3a. In cases (ii)-(iv), it is owed to a faster (with respect to length) broadening of the optical pump pulse. This is because, THz generation initially occurs at a faster rate because of the higher intensity. However, since the generation of THz is necessarily accompanied by spectral broadening of the optical pump pulse, coherent build-up of THz ceases at shorter distances.

In summary, cascading effects in conjunction with GVD-AD represent the strongest limitation to THz generation in lithium niobate for 1 µm pumping. The inclusion of these effects significantly reduces the maximum achievable conversion efficiency. The effective lengths at which the conversion efficiency peaks reduce significantly and are on the order of ~1mm. For such short propagation lengths, absorption plays a minor role in limiting conversion efficiency. This is an additional reason why FCA is not expected to play a major role for pumping at 1µm. For 800nm pumping, in addition to cascading effects and GVD-AD, three-photon absorption may lead to considerable FCA of THz radiation [14].

## 5. Verification using a 2-D model with complete description of beam parameters

In OR using tilted-pulse-fronts, there is a spatial separation of optical frequency components due to angular dispersion. Furthermore, the generated THz propagates in a direction, different from the optical pump. Due to these separating field profiles, a realistic simulation can only be expected from a 2-D spatial model. We developed a comprehensive 2-D spatial model which takes into consideration the non-collinear propagation of the optical and THz fields and their mutual interaction. It includes angular dispersion, cascading effects as well as SPM and SRS effects. The crystal geometry and reflections of THz radiation at the crystal boundary are also considered. The formulation of this model is a 2-D extension of the earlier presented 1-D model and will be presented in detail elsewhere.

We use this 2-D model to confirm the conclusion that cascading effects in conjunction with GVD-AD is the strongest limitation to the scaling of conversion efficiency. The simulation parameters are the same as those used in Fig.3a of Section 5. An optical pump beam with input radius ($1/e^2$) of 2.5mm was incident on the diffraction grating. The transformation of the beam by optical elements was computed using dispersive ray pulse matrices [23]. It impinged the crystal at a distance 2.5mm from the apex of the crystal. Material dispersion and THz absorption are considered in all cases. The simulation results are shown in Fig.5 (a) and (b). The area contained within the box delineated with green lines in Fig.5, represents regions with a non-zero $\chi_{eff}^{(2)}$. The right edge of the computational space represents the output facet of the crystal from which the THz radiation emerges.

This choice of co-ordinate axis is the same as that depicted in Fig.1. The oblique beam (cyan/light-blue colored) depicts the optical pump fluence $\propto \int_0^\infty |E_{op}(\omega,x,z)|^2 \, d\omega$. Here x is the transverse direction, perpendicular to the direction of THz propagation. The optical pump pulse enters the crystal normal to the input surface as seen in Fig.5. In Fig.5, the angular dispersion of the optical pump beam cannot be seen with the naked eye because even for a large bandwidth of 100nm, the beam is still relatively paraxial (with $k_x \ll k_z$). However, the evidence of angular dispersion is in the generation of THz. Without the appropriate angular dispersion, phase matching is not possible and THz generation will not occur. The red colored regions represent the total generated THz fluence $\propto \int_0^\infty |E_{THz}(\Omega,x,z)|^2 \, d\Omega$

### (a) SPM, GVD-AD, material dispersion, absorption

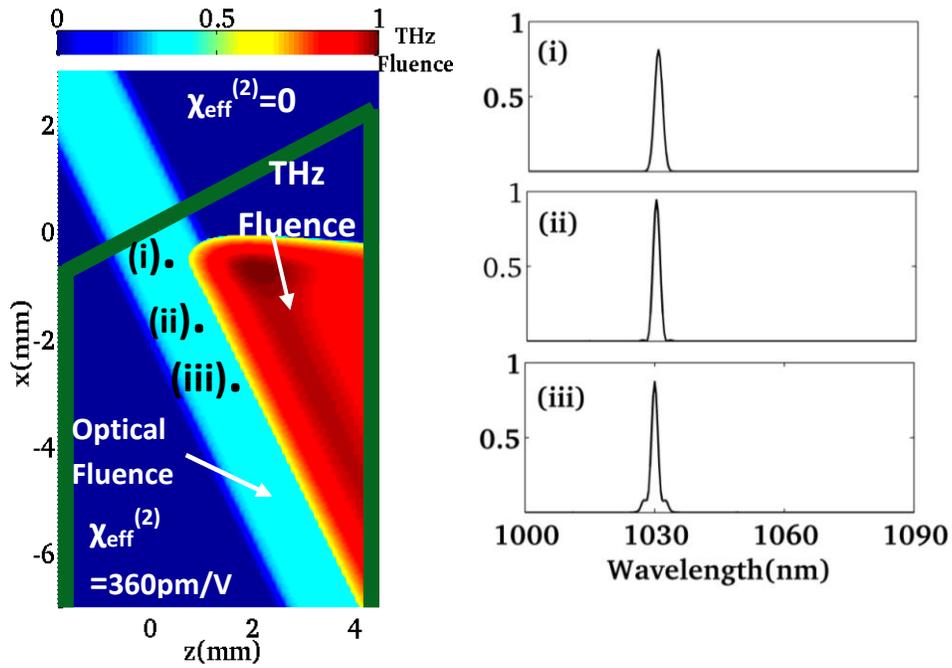

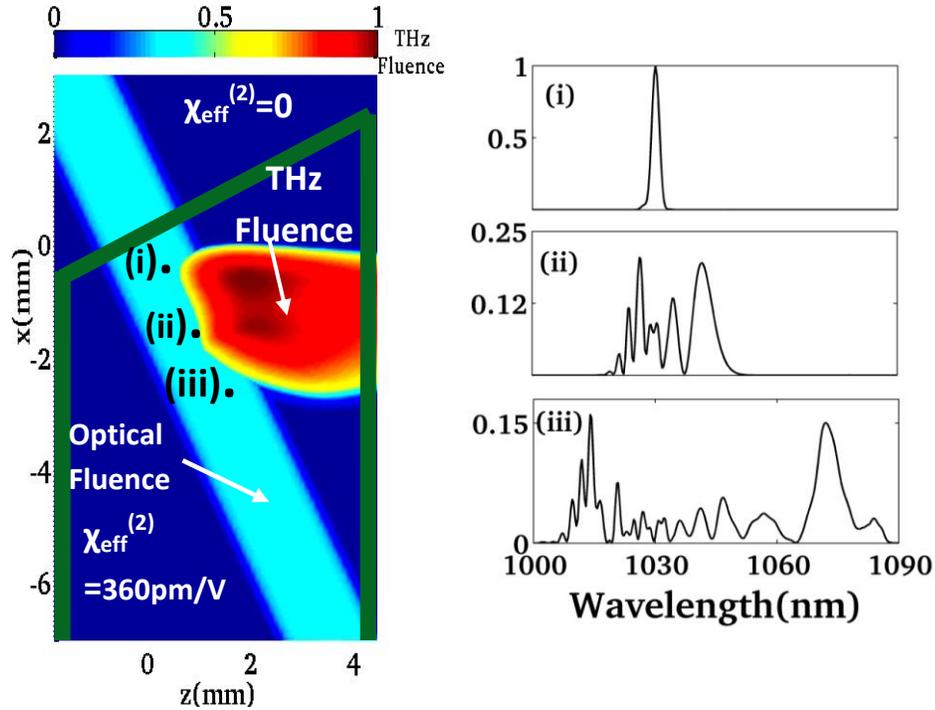

Fig.4: Simulation parameters are the same as in Fig.3. (a) THz fluence (red) and optical fluence (oblique, cyan/light-blue) when only SPM effects are included. Regions within the green box have non-zero $\chi^{(2)}$. THz generation occurs along the full length of optical pump pulse propagation. minimal broadening by SPM is seen between pulse spectra at locations (i) to (iii). A slight narrowing of the spectrum is seen in (ii), compared to(i) due to the spatial chirp associated with angular dispersion. The conversion efficiency is 2.28% (b) THz and optical fluence when cascading effects are included. Note how the pulse spectrum in location (i) is rapidly broadened in locations (ii),(iii). THz generation ceases once the pulse spectrum has drastically broadened. The conversion efficiency is only 0.85%.

In Fig.4a, we calculate the generated THz fluence when only SPM and GVD-AD are included. THz continues to be generated over the entire length of the optical pump pulse propagation. The optical pump pulse spectra at locations delineated by labels (i)-(iii) are shown. No appreciable broadening of the optical pump pulse spectrum due to SPM is seen, which explains why THz continues to be generated, unabated. There is a slight narrowing of the spectrum between locations (i) and (ii). This is a consequence of angular dispersion, which leads to different spectra at different spatial locations, even in the absence of THz generation. This result is in agreement with the experimental observation that no appreciable broadening due to SPM is observed [7],[14] and simulations (Fig.2a). The conversion efficiency in Fig.4a is 2.28%.

In Fig.4b, cascading effects and GVD-AD are considered while SPM effects are excluded and all simulation parameters are kept fixed. The scale for the THz and optical fluences are the same as in Fig.4a. Initially, there is a growth in the amount of generated THz fluence analogous to Fig.3a. However, after a short distance of propagation, the generation of THz ceases to occur. A conversion efficiency of only 0.85% is achieved. The optical spectra at the spatial locations delineated by labels (i)- (iii) are shown. It can be seen that for location (i), prior to appreciable THz being generated, there is no noticeable broadening of the optical pump spectrum. However, in (ii) and (iii), there is significant broadening of the optical pump spectra. After extensive broadening in (iii), no further THz is generated. Therefore, the 2-D model is in agreement with the 1-D model in that cascading effects in conjunction with angular dispersion limit the conversion efficiency more strongly than SPM, SRS. The extent of broadening at location (iii) is larger than the experimental results since spatial averaging effects are not included.

## 6. Conclusion

In conclusion, we presented 1-D and 2-D spatial models which simultaneously consider the nonlinear coupled interaction of THz and optical fields, angular and material dispersion, SPM, SRS and THz absorption. The models were verified by observing that the extent of spectral broadening of the optical pump is commensurate with the amount of THz generated, which is in good agreement with experiments [7]. There is also a good match in the calculated conversion efficiency values which validates our modeling approach. We showed that the large frequency down-shift and spectral broadening due to THz generation (cascading effects) accentuates the phase mismatch caused by GVD-AD in lithium niobate. This turns out to be the strongest limitation to THz generation in lithium niobate for 1μm pumping. Neglecting these cascading effects therefore grossly overestimates the achievable conversion efficiencies. These conclusions were verified with a comprehensive 2-D spatial simulation of OR using tilted-pulse-fronts. Since, sufficient generation of THz can be expected to lead to some amount of spectral broadening of the optical pump pulse , consideration of cascading effects will be important for other OR systems as well. These results have important implications for the maximum conversion efficiencies that can be achieved with such systems and are crucial to the design of high-energy THz sources. Hence, this work paves the way for further discussions on optimization of OR conversion efficiency with high-energy pump sources.

**Acknowledgment:**

The authors acknowledge helpful discussions with Prof. Erich Ippen, Dr. Damian Schimpf, Dr. Kyung-Han Hong and Mr. Patrick Callahan. This work was supported by DARPA under contract N66001-11-1-4192, by the AFOSR under grant AFOSR - A9550-12-1-0499, the Center for Free-Electron Laser Science at DESY and the excellence cluster "The Hamburg Centre for Ultrafast Imaging- Structure, Dynamics and Control of Matter at the Atomic Scale" of the Deutsche Forschungsgemeinschaft. Dr. Wu acknowledges support by a Research Fellowship from the Alexander von Humboldt Foundation.

**References and links**